\documentclass[prl,onecolumn,showpacs,amsmath,amssymb,superscriptaddress]{revtex4}
\usepackage{pifont,amssymb,epsf,graphicx}

\begin{document}
\title{Exact Solution for Three-Dimensional Ising Model}

\author{Degang Zhang}
\affiliation{College of Physics and Electronic Engineering, Sichuan Normal University,
Chengdu 610101, China}
\affiliation{Institute of Solid State Physics, Sichuan Normal
University, Chengdu 610101, China}

\begin{abstract}

Three-dimensional Ising model in zero external field is exactly solved by
operator algebras, similar to the Onsager's approach in two dimensions.
The partition function of the simple cubic crystal imposed by the periodic boundary condition
along two directions and the screw boundary condition along the
third direction  is calculated rigorously. In the thermodynamic limit an integral replaces
a sum in the formula of the partition function. The critical temperatures, at which
order-disorder transitions in the infinite crystal occur along three axis directions, are determined.
The analytical expressions for the internal energy and the specific heat are also presented.

\end{abstract}


\maketitle

\section{I. Introduction}

The exact solution for three-dimensional (3D) Ising model has been one of the greatest challenges
to the physics community for decades. In 1925, Ising presented the simple statistical model
in order to study the order-disorder transition in ferromagnets [1]. Subsequently
the so-called Ising model has been widely applied in condensed matter physics. Unfortunately,
one-dimensional Ising model has no phase transition at nonzero temperature. However,
such systems could have a transition at nonzero temperature in higher dimensions [2].
In 1941, Kramers and Wannier
located the critical point of two-dimensional (2D) Ising model at finite temperature
by employing the dual transformation[3]. About two and a half years later
Onsager solved exactly 2D Ising model by using an algebraic approach [4]
and calculated the thermodynamic properties.
Contrary to the continuous internal energy, the specific heat becomes infinite at
the transition temperature $T=T_c$ given by the condition:
$\sinh\frac{2J}{k_BT_c}\sinh\frac{2J^\prime}{k_BT_c}=1$, where $(J^\prime J)$ are
the interaction energies along two perpendicular directions in a plane, respectively.
Later, the partition function of 2D Ising model was also re-evaluated by a
spinor analysis [5]. Up to now many 2D statistical systems have been
exactly solved [6].

Since Onsager exactly solved 2D Ising model in 1944, much attention has been paid to the
investigation of 3D Ising model.
In Ref. [7], Griffiths presented the first rigorous proof of an order-disorder phase transition
in 3D Ising model at finite temperature by extending the Peierls's argument in 2D case [2]. In 2000,
Istrail proved that solving 3D Ising model on the lattice is an NP-complete problem [8].
We also note that the critical properties of 3D Ising model were widely explored by employing
conformal field theories [9,10,11], self-consistent Ornstein-Zernike approximation [12],
Renormalization group theory [13],
Monte Carlo Simulations [14], the principal components analysis [15], and etc..
However, despite great efforts, 3D Ising model has not been solved exactly yet due to its complexity.
It is out of question that an exact solution of 3D Ising model would be a huge jump forward,
since it can be used to not only describe a broad class of phase transitions ranging from
binary alloys, simple liquids and lattice gases to easy-axis magnets [16], but also verify
the correctness of numerical simulations and finite-size scaling theory in three dimensions.

Because there is no dual transformation, the critical point of 3D Ising model cannot
be fixed by such a symmetry.
We also discover that it is impossible to write out the Hamiltonian along the third
dimension of 3D Ising model with periodic boundary conditions (PBCs) in terms of the Onsager's operators.
In addition, due to the existence of nonlocal rotation, 3D Ising model with PBCs seems not to be
also solved by the spinor analysis [5]. Therefore, the key to solve 3D Ising model
is to find out the operator expression of the interaction along the third dimension.
We note that the transfer matrix in 3D Ising model is constructed by the spin configurations on a plane,
which the boundary conditions (BCs) play an important role to solve exactly 3D Ising model.
In this paper, we introduce a set of operators, which is similar to that in solving
2D Ising model [4]. Under suitable BCs, 3D Ising model with vanishing external
field can be described by the operator algebras, and thus can be solved exactly.

\section{II. Theory}

Consider a simple cubic lattice with $l$ layers, $n$ rows per layer, and $m$ sites
per row. Then the Hamiltonian of 3D Ising model is $H=-\sum^{m,n,l}_{i,j,k=1}
(J_1\sigma^z_{ijk}\sigma^z_{i+1kj}+J_2\sigma^z_{ijk}\sigma^z_{ij+1k}
+J\sigma^z_{ijk}s^z_{ijk+1})$, where
$\sigma^z_{ijk}=\pm 1$ is the spin on the site $[ijk]$. Assume that $\nu_k$ labels the
spin configurations in the $k$th layer, we have $1\leq\nu_k\leq 2^{mn}$.
As a result, the energy of a spin configuration of the crystal $E_{sc}=
\sum^l_{k=1}E_1(\nu_k)+\sum^l_{k=1}E_2(\nu_k)+\sum^l_{k=1}E(\nu_k,\nu_{k+1})$,
where $E_1(\nu_k)$ and $E_2(\nu_k)$ are the energies along two perpendicular
directions in the $k$th layer, respectively, and $E(\nu_k,\nu_{k+1})$ is the
energy between two adjacent layers. Now we define $(V_1V_2)_{\nu_k\nu_k}
=\sum_{\nu^\prime_k}(V_1)_{\nu_k\nu^\prime_k}(V_2)_{\nu^\prime_k\nu_k}
=(V_1)_{\nu_k\nu_k}(V_2)_{\nu_k\nu_k}
\equiv \exp[-E_1(\nu_k)/(k_BT)]\times\exp[-E_2(\nu_k)/(k_BT)]$ and
$(V_3)_{\nu_k\nu_{k+1}}\equiv \exp[-E(\nu_k,\nu_{k+1})/(k_BT)]$.
Here we use the periodic boundary conditions along both $(0 1 0)$ and
$(0 0 1)$ directions and the screw boundary condition along the
$(1 0 0)$ direction for simplicity [3] (see Fig. 1).
So the spin configurations along the ${\bf X}$ direction in a layer can be
described by the spin variables $\sigma^z_1, \sigma^z_2,\cdots, \sigma^z_{mn}$.
Because the probability of a spin configuration is proportional to
$\exp[-E_{sc}/(k_BT)]=(V_1V_2)_{\nu_1\nu_1}(V_3)_{\nu_1\nu_2}
(V_1V_2)_{\nu_2\nu_2}(V_3)_{\nu_2\nu_3}\cdots
(V_1V_2)_{\nu_l\nu_l}(V_3)_{\nu_l\nu_1}$, the partition function of
3D Ising model is
$$\begin{array}{lll}
Z&=&\sum_{\nu_1,\nu_2,\cdots,\nu_l}
(V_1V_2)_{\nu_1\nu_1}(V_3)_{\nu_1\nu_2}\cdots
(V_1V_2)_{\nu_l\nu_l}(V_3)_{\nu_l\nu_1}\\
&\equiv& {\rm tr}(V_1V_2V_3)^l.
\end{array}\eqno{(1)}$$
We note that $V_1$, $V_2$ and $V_3$ are $2^{mn}$-dimensional matrices,
and both  $V_1$ and $V_2$  are diagonal. Following Ref. [4], we obtain
$$\begin{array}{lll}
V_1&=&\exp(H_1\sum^{mn}_{\tau=1}\sigma^z_\tau\sigma^z_{\tau+1})\equiv\exp(H_1H_x),\\
V_2&=&\exp(H_2\sum^{mn}_{\tau=1}\sigma^z_\tau\sigma^z_{\tau+m})\equiv\exp(H_2H_y),\\
V_3&=&[2\sinh(2H)]^{mn/2}\exp(H^*\sum^{mn}_{\tau=1}\sigma^x_\tau)\\
&\equiv &[2\sinh(2H)]^{mn/2}\exp(H^*H_z),
\end{array}\eqno{(2)}$$
where $H_1=J_1/(k_BT), H_2=J_2/(k_BT), H=J/(k_BT)$, and $H^*=\frac{1}{2}
\ln\coth H=\tanh^{-1}(e^{-2H})$.

\begin{figure}
\rotatebox[origin=c]{0}{\includegraphics[angle=0,
           height=2.5in]{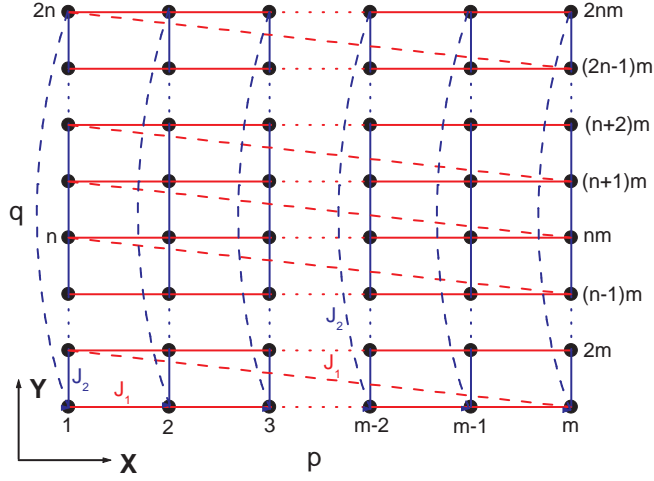}}

\caption {(Color online) The lattice structure in each layer of the simple cubic crystal.}
\end{figure}

In order to diagonalize the transfer matrix $V\equiv V_1V_2V_3$, following
the Onsager's famous work in two dimensions, we first introduce the operators
$$L_{a,a}=-\sigma^x_a,~~~
L_{a,b}=\sigma^z_a\sigma^x_{a+1}\sigma^x_{a+2}\cdots \sigma^x_{b-1}\sigma^z_b \eqno{(3)}$$
in spin space $\Gamma $
along the {\bf X} direction under the boundary conditions mentioned above.
Here $a, b=1, 2, \cdots, 2mn$, $\sigma^x_a$, $\sigma^y_a$ and $\sigma^z_a$ are the
Pauli matrices at site $a$, respectively. Then we have $L^2_{a,b}=1$ and
$$L_{a,b+mn}=L_{a+mn,b}=-QL_{a,b}=-L_{a,b}Q\eqno{(4)}$$
with $Q\equiv \prod^{nm}_{a=1}\sigma^x_a=\pm 1$.
It is obvious that the period of $L_{a,b}$ is 2mn. We note that
these operators $L_{a,b}$ are identical to $P_{ab}$ in Ref. [4] except $mn$ replaces $n$.

$H_x$ and $H_z$ in the transfer matrix $V$ can be expressed as
$$\begin{array}{lll}
H_x=\sum^{mn}_{a=1}L_{a,a+1},\\
H_z=\sum^{mn}_{a=1}\sigma^x_a=-\sum^{mn}_{a=1}L_{a,a}.
\end{array}\eqno{(5)}$$

Following Onsager's idea [4], we introduce the operators
$$\begin{array}{lll}
\alpha_r&=&-\frac{1}{4mn}\sum^{2mn}_{a,b=1}L_{a,b}\cos\frac{(a-b)r\pi}{mn},\\
\beta_r&=&-\frac{1}{4mn}\sum^{2mn}_{a,b=1}L_{a,b}\sin\frac{(a-b)r\pi}{mn},\\
\gamma_r&=&\frac{i}{8mn}\sum^{2mn}_{a,b=1}(L_{a,x}L_{b,x}-L_{x,a}L_{x,b})\sin\frac{(a-b)r\pi}{mn}
\end{array}\eqno{(6)} $$
where $x$ is an arbitrary index. Obviously, we have
$\alpha_{-r}=\alpha_r$,
$\beta_{-r}=-\beta_r$, $\beta_0=\beta_{mn}=0$,
$\gamma_{-r}=-\gamma_r$, and $\gamma_0=\gamma_{mn}=0$.
Eqs. (6) can be rewritten as
$$\begin{array}{lll}
\alpha_r&=&-\frac{1}{2mn}\sum^{2mn}_{s=1}A_s \cos\frac{rs\pi}{mn},\\
\beta_r&=&\frac{1}{2mn}\sum^{2mn}_{s=1}A_s \sin\frac{rs\pi}{mn},\\
\gamma_r&=&-\frac{i}{2mn}\sum^{2mn}_{s=1}G_s \sin\frac{rs\pi}{mn},
\end{array}\eqno{(7)}$$
where $A_s=\sum^{mn}_{a=1}L_{a,a+s}$ and
$G_s=\frac{1}{2}\sum^{mn}_{a=1}(L_{a,x}L_{a+s,x}-L_{x,a}L_{x,a+s})$.
According to the orthogonal properties of the coefficients, we obtain
$$\begin{array}{lll}
A_s&=&\sum^{2mn}_{r=1}[-\alpha_r \cos\frac{rs\pi}{mn}+\beta_r \sin\frac{rs\pi}{mn}],\\
G_s&=&i\sum^{2mn}_{r=1}\gamma_r \sin\frac{rs\pi}{mn}.
\end{array}\eqno{(8)}$$

From Eqs. (5)-(8), $H_x$ and $H_z$  have the expansions
$$\begin{array}{l}
H_x=A_1=-2\sum^{mn-1}_{r=1}(\alpha_r\cos\frac{r\pi}{mn}
          -\beta_r\sin\frac{r\pi}{mn})\\
          ~~~~~~~~~-\alpha_0+\alpha_{mn},\\
H_z=-A_0=\alpha_0+2\sum^{mn-1}_{r=1}\alpha_r+\alpha_{mn}.
\end{array}\eqno{(9)}$$
Because $A_{mn+s}=-QA_s=-A_sQ$ and $G_{mn+s}=-QG_s=-G_sQ$,
and combining with Eqs. (8), we have
$$[1+(-1)^rQ]\alpha_r=[1+(-1)^rQ]\beta_r=[1+(-1)^rQ]\gamma_r=0.\eqno{(10)}$$
When $Q=1$, $\alpha_{2r}=\beta_{2r}=\gamma_{2r}=0$ while
$\alpha_{2r+1}=\beta_{2r+1}=\gamma_{2r+1}=0$ if $Q=-1$.
So we can investigate the algebra (8) with $Q=1$ or -1 independently. However,
we keep them together for convenience. In order to diagonalize the transfer
matrix $V$, we must first determine the commutation relations among the operators $\alpha_r$,
$\beta_r$ and $\gamma_r$. Similar to those calculations in Ref. [4],
we obtain
$$\begin{array}{l}
[A_i, A_j]=4G_{i-j}, ~~~[G_i, G_j]=0,\\

[G_i, A_j]=2(A_{j+i}-A_{j-i}).
\end{array}
\eqno{(11)}$$
Substituting Eqs. (8) into Eqs. (11), we arrive at
$$[\alpha_r, \beta_r]=2i\gamma_r,~~[\beta_r, \gamma_r]=2i\alpha_r,~~
[\gamma_r, \alpha_r]=2i\beta_r,
\eqno{(12)}$$
where $r=1, 2, \cdots, mn-1$, and all the other commutators vanish.
Obviously, the algbra (12) is associated with the site $r$, and hence is local.
Because $\alpha_r$, $\beta_r$, and $\gamma_r$ obey the same commutation relations
with $-X_r$, $-Y_r$, and $-Z_r$ in Ref. [4], we have the further relations
$$\begin{array}{l}
\alpha^2_0=\frac{1}{2}(1-Q)=R_0,\\
\alpha^2_{mn}=\frac{1}{2}[1-(-1)^{mn}Q]=R_{mn},\\
\alpha_r\beta_r=i\gamma_r,~~ \beta_r\gamma_r=i\alpha_r,~~ \gamma_r\alpha_r=i\beta_r,\\
\alpha^2_r=\beta^2_r=\gamma^2_r=R^2_r=R_r,
~~\alpha_r=R_r\alpha_r=\alpha_r R_r,\\
\beta_r=R_r\beta_r=\beta_r R_r,
\gamma_r=R_r\gamma_r=\gamma_r R_r.
\end{array}\eqno{(13)}$$

We note that $A_{sm}=\sum_{p=1}^{m}A_{p,s}=\sum_{p=1}^{m}\sum_{a=1}^{n}L^p_{a,a+s}
=\sum_{p=1}^{m}\sum_{a=1}^{n}L_{p+(a-1)m,p+(a-1+s)m}$
and $G_{sm}=\sum_{p=1}^{m}G_{p,s}=\frac{1}{2}\sum_{p=1}^{m}\sum_{a=1}^{n}[
L_{p+(a-1)m,x} \times L_{p+(a-1+s)m,x}-L_{x,p+(a-1)m}L_{x,p+(a-1+s)m}]$,
where $s=1, 2, \cdots, 2n$, and
$$\begin{array}{lll}
A_{p,s}&=&\sum^{2n}_{q=1}\{-\alpha_{p+(q-1)m} \cos\frac{[p+(q-1)m]s\pi}{n}\\
 & &+\beta_{p+(q-1)m} \sin\frac{[p+(q-1)m]s\pi}{n}\},\\
G_{p,s}&=&i\sum^{2n}_{q=1}\gamma_{p+(q-1)m} \sin\frac{[p+(q-1)m]s\pi}{n}.
\end{array}\eqno{(14)}$$
When $m=p=1$, Eqs. (14) recover the results in two dimensions [4].
It is obvious that $A_{p,i}$ and $G_{p,j}$ also satisfy the
commutation relations (11). When $p\not =p^\prime$, $[A_{p,i}, A_{p^\prime,i^\prime}]=
[G_{p,j}, G_{p^\prime,j^\prime}]=[A_{p,i}, G_{p^\prime,i^\prime}]=0$.

We have obtained the expressions of $H_x$ and $H_z$ in terms of
the operators $\alpha_r$, $\beta_r$ and $\gamma_r$ in the space $\Gamma$. In order to get the
Hamiltonian in the third dimension, we project the operator algebra in
the space $\Gamma$ into the ${\bf Y}$ direction. Then we have $m$ subspaces
$\Gamma_p (p=1, 2, \cdots, m)$, in which the operator algebra with period $2n$
is same with that in $\Gamma$. In $\Gamma_p$, we define
$$\begin{array}{l}
{\cal L}^p_{a,a}=-\sigma^x_{p+(a-1)m},\\
{\cal L}^p_{a,b}=\sigma^z_{p+(a-1)m}\sigma^x_{p+am}
\cdots \sigma^x_{p+(b-2)m}\sigma^z_{p+(b-1)m}\end{array} \eqno{(15)}$$
along the ${\bf Y}$ direction. Then we have ${\cal A}_{p,s}=\sum_{a=1}^{n}{\cal L}^p_{a,a+s}$
and ${\cal G}_{p,s}=\frac{1}{2}\sum_{a=1}^{n}[
{\cal L}_{p+(a-1)m,x}{\cal L}_{p+(a-1+s)m,x}
-{\cal L}_{x,p+(a-1)m}\times {\cal L}_{x,p+(a-1+s)m}]$, which also obey
the same commutation relations (11) and (12), similar to $A_{p,s}$ and $G_{p,s}$.
Then the Hamiltonian $H_y=\sum^{m}_{p=1}{\cal A}_{p,1}$.

\begin{figure}
\rotatebox[origin=c]{0}{\includegraphics[angle=0,
           height=2.1in]{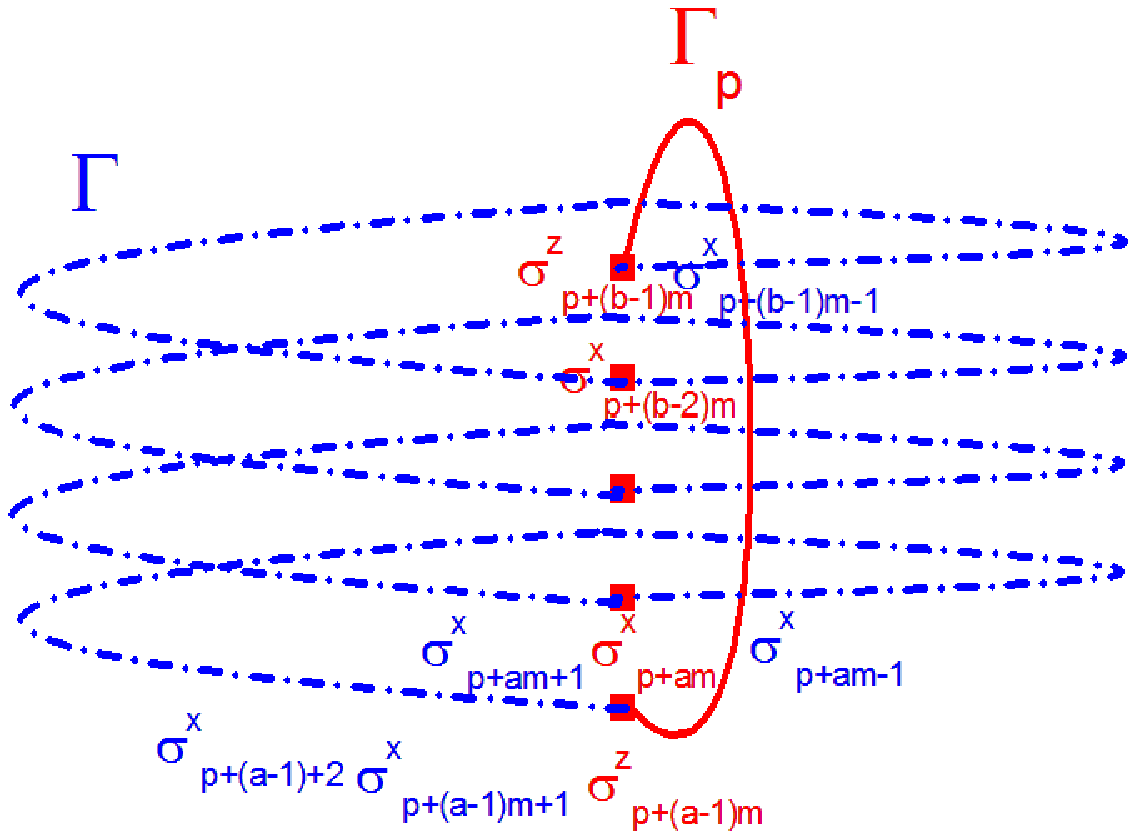}}

\caption {(Color online) Operator renormalization: schematic of ${\cal L}^p_{a,b}$ in $\Gamma_p$
along the ${\bf Y}$ direction and $L^p_{a,b}$ in $\Gamma$ along the ${\bf X}$ direction.}
\end{figure}

Because $[{\cal L}^p_{a,a+s}, L^p_{b,b+s}]=0$ (see Fig. 2),
we have $[{\cal A}_{p,s}, A_{p,s}]=0$, which leads to
${\cal A}_{p,s}\equiv A_{p,s}$ due to their common local algebra (12).
{\it This is a renormalization of operator, which means that ${\cal A}_{p,s}$ and  $A_{p,s}$ have same eigenfunctions and
eigenvalues in $\Gamma_p$ or $\Gamma$ space}.
We note that $V_2$ is the transfer matrix along ${\bf Y}$ direction, which must be calculated
in $\Gamma$ rather than $\Gamma_p$ space by mapping ${\cal A}_{p,1}\equiv A_{p,1}$ in order to diagonalize total transfer matrix $V$.
Therefore, we have
$$\begin{array}{lll}
H_y&=&\sum^{m}_{p=1}{\cal A}_{p,1}=\sum^{m}_{p=1}A_{p,1}\equiv A_m\\
&=&-\alpha_0-2\sum^{mn-1}_{r=1}(\alpha_r\cos\frac{r\pi}{n}
          -\beta_r\sin\frac{r\pi}{n})\\
 &&         -(-1)^{m}\alpha_{mn}.
\end{array}\eqno{(16)}$$
Here, we would like to mention that $H_z=-\sum^{m}_{p=1}{\cal A}_{p,0}\equiv -A_0$,
which is same with that in (9). This means that when $J_1=0$, the Hamiltonian of
2D Ising model is recovered immediately.

Because $[Q, H_x]=[Q, H_y]=[Q, H_z]=[Q, V]=0$, $V$ and $Q$ can be simultaneously
diagonalized on the same basis. In other words, the eigenvalue problem of $V$
can be classified by the value $\pm 1$ of $Q$.

The transfer matrix $V$ with Eqs. (9) and (16) becomes
$$\begin{array}{lll}
V&=&[2\sinh(2H)]^{\frac{mn}{2}}e^{H_1A_1}e^{H_2A_m}e^{-H^*A_0}\\
&=&[2\sinh(2H)]^{\frac{mn}{2}}e^{(H^*-H_1-H_2)\alpha_0}\\
&&\times\prod^{mn-1}_{r=1}U_re^{[H^*+H_1-(-1)^{m}H_2]\alpha_{mn}},
\end{array}\eqno{(17)}$$
where
$$\begin{array}{lll}
U_r&=&e^{-2H_1(\alpha_r \cos\frac{r\pi}{mn}-\beta_r \sin\frac{r\pi}{mn})}\\
   &&\times e^{-2H_2(\alpha_r \cos\frac{r\pi}{n}-\beta_r \sin\frac{r\pi}{n})}
      e^{2H^*\alpha_r}.
\end{array}$$

In order to obtain the eigenvalues of the transfer matrix $V$, we first diagonalize $U_r$
by employing the general unitary transformation:
$$\begin{array}{l}
e^{\frac{i}{2}\eta_r\gamma_r}e^{a_r(\alpha_r\cos\theta_r+\beta_r\sin\theta_r)}U_r\\
\times e^{-a_r(\alpha_r\cos\theta_r+\beta_r\sin\theta_r)}
e^{-\frac{i}{2}\eta_r\gamma_r}=e^{\xi_r\alpha_r}.
\end{array}\eqno{(18)}$$
Here  $\theta_r$ is an arbitrary constant and can be taken to be zero
without loss of generality, and
$$\begin{array}{l}
\cosh\xi_r={\cal D}_r,\\
\sinh\xi_r\cos\eta_r={\cal A}_r,~~\tanh(2a_r)=\frac{{\cal C}_r}{{\cal B}_r},\\
\sinh\xi_r\sin\eta_r={\cal B}_r\cosh(2a_r)-{\cal C}_r\sinh(2a_r),\\
\end{array}\eqno{(19)}$$
where
$$\begin{array}{lll}
{\cal A}_r&=&\cosh(2H_1)\cosh(2H_2)\sinh(2H^*)\\
&&      -\sinh(2H_1)\cosh(2H_2)\cosh(2H^*)\cos\frac{r\pi}{mn}\\
&&      -\cosh(2H_1)\sinh(2H_2)\cosh(2H^*)\cos\frac{r\pi}{n}\\
&&      +\sinh(2H_1)\sinh(2H_2)\sinh(2H^*)\cos\frac{(m-1)r\pi}{mn},\\
{\cal B}_r&=&\sinh(2H_1)\cosh(2H_2)\cosh(2H^*)\sin\frac{r\pi}{mn}\\
&&      +\cosh(2H_1)\sinh(2H_2)\cosh(2H^*)\sin\frac{r\pi}{n}\\
&&      +\sinh(2H_1)\sinh(2H_2)\sinh(2H^*)\sin\frac{(m-1)r\pi}{mn},\\
{\cal C}_r&=&\sinh(2H_1)\cosh(2H_2)\sinh(2H^*)\sin\frac{r\pi}{mn}\\
&&      +\cosh(2H_1)\sinh(2H_2)\sinh(2H^*)\sin\frac{r\pi}{n}\\
&&      +\sinh(2H_1)\sinh(2H_2)\cosh(2H^*)\sin\frac{(m-1)r\pi}{mn},\\
{\cal D}_r&=&\cosh(2H_1)\cosh(2H_2)\cosh(2H^*)\\
&&      -\sinh(2H_1)\cosh(2H_2)\sinh(2H^*)\cos\frac{r\pi}{mn}\\
&&      -\cosh(2H_1)\sinh(2H_2)\sinh(2H^*)\cos\frac{r\pi}{n}\\
&&      +\sinh(2H_1)\sinh(2H_2)\cosh(2H^*)\cos\frac{(m-1)r\pi}{mn}.
\end{array}$$
We note that ${\cal D}^2_r+{\cal C}^2_r-{\cal A}^2_r-{\cal B}^2_r\equiv 1$,
which ensures that 3D Ising model can be solved exactly
in the whole parameter space. When $H_2=0 (i.e. J_2=0)$ and $n=1$, or
$H_1=0 (i.e. J_1=0)$ and $m=1$, we have $a_r=H^*$. So Eqs. (19) recover
the Onsager's results in 2D Ising model [4].

Then the transfer matrix $V$ has a diagonal form
$$\begin{array}{l}
e^{\sum^{mn-1}_{r=1}\frac{i}{2}\eta_r\gamma_r}e^{\sum^{mn-1}_{r=1}a_r\alpha_r}V
e^{-\sum^{mn-1}_{r=1}a_r\alpha_r}\\
\times e^{-\sum^{mn-1}_{r=1}\frac{i}{2}\eta_r\gamma_r}=
[2\sinh(2H^*)]^{\frac{mn}{2}}\\
\times e^{(H^*-H_1-H_2)\alpha_0+\sum^{mn-1}_{r=1}\xi_r\alpha_r
+[H^*+H_1-(-1)^{m}H_2]\alpha_{mn}}.
\end{array}\eqno{(20)}$$

\section{III. Transformations}

\subsection{A. Transformation 1}

In order to explore the symmetries in 3D Ising model,
we take the transformation
$$\begin{array}{lll}
\alpha^*_r&=&-\alpha_r\cos\frac{r\pi}{mn}+\beta_r\sin\frac{r\pi}{mn},\\
\beta^*_r&=&\alpha_r\sin\frac{r\pi}{mn}+\beta_r\cos\frac{r\pi}{mn},~~
\gamma^*_r=-\gamma_r.
\end{array}\eqno{(21)}$$
It is easy to prove that $\alpha^*_r$, $\beta^*_r$ and $\gamma^*_r$ satisfy the same
commutation relations with $\alpha_r$, $\beta_r$ and $\gamma_r$.
Then we have
$$\begin{array}{l}
H_x=\alpha^*_0+2\sum^{mn-1}_{r=1}\alpha^*_r+\alpha^*_{mn},\\
H_y=\alpha^*_0+2\sum^{mn-1}_{r=1}[\alpha^*_r\cos\frac{(m-1)r\pi}{mn}
          +\beta^*_r\sin\frac{(m-1)r\pi}{mn}]\\
    ~~~~~~~~-(-1)^{m}\alpha^*_{mn},\\
H_z=-\alpha^*_0-2\sum^{mn-1}_{r=1}[\alpha^*_r\cos\frac{r\pi}{mn}
          -\beta^*_r\sin\frac{r\pi}{mn}]+\alpha^*_{mn}.
\end{array}\eqno{(22)}$$
Obviously, by comparing with Eqs. (9) and (16), such a transformation (21) exchanges the interaction forms in
$(1, 0, 0)$ and $(0, 0, 1)$ directions (i.e. $H_x$ and $H_z$), but changes the interaction form
in $(0, 1, 0)$ direction (i.e. $H_y$). Therefore, 3D Ising model has no a dual transformation,
and the critical point cannot be fixed by the Kramers and Wannier's approach [3].

The transfer matrix can be expressed as
$$\begin{array}{lll}
V&=&[2\sinh(2H)]^{\frac{mn}{2}}e^{H_1A_1}e^{H_2A_m}e^{-H^*A_0}\\
&=&[2\sinh(2H)]^{\frac{mn}{2}}e^{(H_1+H_2-H^*)\alpha^*_0}\\
&&\times\prod^{mn-1}_{r=1}U^*_re^{[H_1-(-1)^{m}H_2+H^*]\alpha^*_{mn}},
\end{array}\eqno{(23)}$$
where
$$\begin{array}{lll}
U^*_r&=&e^{2H_1\alpha^*_r}
e^{2H_2[\alpha^*_r \cos\frac{(m-1)r\pi}{mn}+\beta^*_r \sin\frac{(m-1)r\pi}{mn}]}\\
 &&\times e^{-2H^*(\alpha^*_r \cos\frac{r\pi}{mn}-\beta^*_r \sin\frac{r\pi}{mn})}.
\end{array}$$

Following the procedure above, we can diagonalize the transfer matrix $V$, i.e.
$$\begin{array}{l}
e^{\sum^{mn-1}_{r=1}\frac{i}{2}\eta^*_r\gamma^*_r}e^{\sum^{mn-1}_{r=1}a^*_r\alpha^*_r}V
e^{-\sum^{mn-1}_{r=1}a^*_r\alpha^*_r}\\
\times e^{-\sum^{mn-1}_{r=1}\frac{i}{2}\eta^*_r\gamma^*_r}=
[2\sinh(2H^*)]^{\frac{mn}{2}}\\
\times e^{(H_1+H_2-H^*)\alpha^*_0+\sum^{mn-1}_{r=1}\xi_r\alpha^*_r
+[H_1-(-1)^{m}H_2+H^*]\alpha^*_{mn}},
\end{array}\eqno{(24)}$$
where
$$\begin{array}{l}
\sinh\xi_r\cos\eta^*_r={\cal A}^*_r,~~\tanh(2a^*_r)=-\frac{{\cal C}_r}{{\cal B}^*_r},\\
\sinh\xi_r\sin\eta^*_r={\cal B}^*_r\cosh(2a^*_r)+{\cal C}_r\sinh(2a^*_r),\\
\end{array}\eqno{(25)}$$
and
$$\begin{array}{lll}
{\cal A}^*_r&=&\sinh(2H_1)\cosh(2H_2)\cosh(2H^*)\\
&&    -\cosh(2H_1)\cosh(2H_2)\sinh(2H^*)\cos\frac{r\pi}{mn}\\
&&      -\sinh(2H_1)\sinh(2H_2)\sinh(2H^*)\cos\frac{r\pi}{n}\\
&&      +\cosh(2H_1)\sinh(2H_2)\cosh(2H^*)\cos\frac{(m-1)r\pi}{mn},\\
{\cal B}^*_r&=&\cosh(2H_1)\cosh(2H_2)\sinh(2H^*)\sin\frac{r\pi}{mn}\\
&&      +\sinh(2H_1)\sinh(2H_2)\sinh(2H^*)\sin\frac{r\pi}{n}\\
&&      +\cosh(2H_1)\sinh(2H_2)\cosh(2H^*)\sin\frac{(m-1)r\pi}{mn}.
\end{array}$$
We also have ${\cal D}^2_r+{\cal C}^2_r-{\cal A}^{*2}_r-{\cal B}^{*2}_r\equiv 1.$

\subsection{B. Transformation 2}

Let
$$\begin{array}{lll}
\alpha^\prime_r&=&-\alpha_r\cos\frac{r\pi}{n}+\beta_r\sin\frac{r\pi}{n},\\
\beta^\prime_r&=&\alpha_r\sin\frac{r\pi}{n}+\beta_r\cos\frac{r\pi}{n},~~
\gamma^\prime_r=-\gamma_r,
\end{array}\eqno{(26)}$$
then we have
$$\begin{array}{l}
H_x=\alpha^\prime_0+2\sum^{mn-1}_{r=1}[\alpha^\prime_r\cos\frac{(m-1)r\pi}{mn}
          -\beta^\prime_r\sin\frac{(m-1)r\pi}{mn}]\\
~~~~~~~~        -(-1)^{m}\alpha^\prime_{mn},\\
H_y=\alpha^\prime_0+2\sum^{mn-1}_{r=1}\alpha^\prime_r+\alpha^\prime_{mn},\\
H_z=-\alpha^\prime_0-2\sum^{mn-1}_{r=1}[\alpha^\prime_r\cos\frac{r\pi}{n}
          -\beta^\prime_r\sin\frac{r\pi}{n}]\\
~~~~~~~~     -(-1)^{m}\alpha^\prime_{mn}.
\end{array}\eqno{(27)}$$
By also comparing with Eqs. (9) and (16), the transformation (26) exchanges the interaction forms in
$(0, 1, 0)$ and $(0, 0, 1)$ directions (i.e. $H_y$ and $H_z$), but changes the interaction form
in $(1, 0, 0)$ direction (i.e. $H_x$). Therefore, such the transformation is not a dual transformation yet,
which cannot be used to fix the critical point [3].

The transfer matrix reads
$$\begin{array}{lll}
V&=&[2\sinh(2H)]^{\frac{mn}{2}}e^{H_1A_1}e^{H_2A_m}e^{-H^*A_0}\\
&=&[2\sinh(2H)]^{\frac{mn}{2}}e^{(H_1+H_2-H^*)\alpha^\prime_0}\\
&&\times\prod^{mn-1}_{r=1}U^\prime_re^{[-(-1)^mH_1+H_2-(-1)^{m}H^*]\alpha^\prime_{mn}},
\end{array}\eqno{(28)}$$
where
$$\begin{array}{lll}
U^\prime_r&=&e^{2H_1[\alpha^\prime_r \cos\frac{(m-1)r\pi}{mn}-\beta^\prime_r \sin\frac{(m-1)r\pi}{mn}]}
        e^{2H_2\alpha^\prime_r}\\
&& \times e^{-2H^*(\alpha^\prime_r \cos\frac{r\pi}{n}-\beta^\prime_r \sin\frac{r\pi}{n})}.
\end{array}$$

Similarly, we have
$$\begin{array}{l}
e^{\sum^{mn-1}_{r=1}\frac{i}{2}\eta^\prime_r\gamma^\prime_r}e^{\sum^{mn-1}_{r=1}a^\prime_r\alpha^\prime_r}V
e^{-\sum^{mn-1}_{r=1}a^\prime_r\alpha^\prime_r}\\
\times e^{-\sum^{mn-1}_{r=1}\frac{i}{2}\eta^\prime_r\gamma^\prime_r}=
[2\sinh(2H^*)]^{\frac{mn}{2}}\\
\times e^{(H_1+H_2-H^*)\alpha^\prime_0+\sum^{mn-1}_{r=1}\xi_r\alpha^\prime_r
+[H_2-(-1)^{m}(H_1+H^*)]\alpha^\prime_{mn}}.
\end{array}\eqno{(29)}$$
Here,
$$\begin{array}{l}
\sinh\xi_r\cos\eta^\prime_r={\cal A}^\prime_r,~~\tanh(2a^\prime_r)=-\frac{{\cal C}_r}{{\cal B}^\prime_r},\\
\sinh\xi_r\sin\eta^\prime_r={\cal B}^\prime_r\cosh(2a^\prime_r)+{\cal C}_r\sinh(2a^\prime_r),\\
\end{array}\eqno{(30)}$$
and
$$\begin{array}{lll}
A^\prime_r&=&\cosh(2H_1)\sinh(2H_2)\cosh(2H^*)\\
&&      -\cosh(2H_1)\cosh(2H_2)\sinh(2H^*)\cos\frac{r\pi}{n}\\
&&      -\sinh(2H_1)\sinh(2H_2)\sinh(2H^*)\cos\frac{(2m-1)r\pi}{mn}\\
&&      +\sinh(2H_1)\cosh(2H_2)\cosh(2H^*)\cos\frac{(m-1)r\pi}{mn},\\
B^\prime_r&=&\cosh(2H_1)\cosh(2H_2)\sinh(2H^*)\sin\frac{r\pi}{n}\\
&&      -\sinh(2H_1)\cosh(2H_2)\cosh(2H^*)\sin\frac{(m-1)r\pi}{mn}\\
&&      +\sinh(2H_1)\sinh(2H_2)\sinh(2H^*)\sin\frac{(2m-1)r\pi}{mn}.
\end{array}$$
The identity ${\cal D}^2_r+{\cal C}^2_r-{\cal A}^{\prime 2}_r-{\cal B}^{\prime 2}_r\equiv 1$
also holds.

\section{IV. Results}

Because $\alpha_0, \alpha_1, \cdots, \alpha_{mn}$ have the common eigenvectors
$\chi_0$ with the corresponding eigenvalues $\Delta_0, \Delta_1, \cdots, \Delta_{mn}$,
from Eq. (20), we have $V\psi=\lambda\psi$, where
$$\begin{array}{rl}
\psi=&e^{-\sum^{mn-1}_{r=1}a_r\alpha_r}
e^{-\sum^{mn-1}_{r=1}\frac{i}{2}\eta_r\gamma_r}\chi_0,\\
\ln\lambda=&\frac{1}{2}mn\ln[2\sinh(2H)]+(H^*-H_1-H_2)\Delta_0\\
&     +\sum^{mn-1}_{r=1}\xi_r\Delta_r+[H_1-(-1)^mH_2+H^*]\Delta_{mn}.
\end{array}\eqno{(31)}$$
At the critical point, we have $\xi_0=H^*-H_1-H_2=0$ [4]. This leads to a critical
temperature $T=T_c$ given by the condition
$$\sinh(2H)\sinh(2H_1+2H_2)=1.\eqno{(32)}$$
If $H_2=0$ or $H_1=0$, we obtain the critical temperature in 2D Ising model [3, 4].
We note that the exact critical line (32) between the ferromagnetic and paramagnetic
phases coincides completely with the result found in the domain wall analysis [17].
In the anisotropic limit, i.e. $\eta=(H_1+H_2)/H\rightarrow 0$, the critical temperature
determined by Eq. (32) also agrees perfectly with the asymptotically exact value
$H=2[{\rm ln}\eta^{-1}-{\rm lnln}\eta^{-1}+0(1)]^{-1}$ shown in Refs. [18,19].

When $H_1=H_2=H$, the critical value $H_c=J/(k_BT_c)=0.30468893$,
which is larger than the conjectured value about 0.2216546 from the previous
numerical simulations [12,14]. We shall see from the analytical expressions (35) and (36) of
the partition function per atom below that this discrepancy mainly comes from the
oscillatory terms with respect to the system size $m$ along {\bf X} direction,
which were not taken into account in all the previous numerical simulations.

We note that the thermodynamic properties of a large crystal are determined by
the largest eigenvalue $\lambda_{\max}$ of the transfer matrix $V$.
Following Ref. [4], we have
$$\begin{array}{l}
\ln\lambda_{\max}-\frac{1}{2}mn\ln[2\sinh(2H)]\\
=\{\begin{array}{l}
\xi_1+\xi_3+\cdots +\xi_{2L-1} ~~{\rm for}~ mn=2L;\\
\xi_1+\xi_3+\cdots +\xi_{2L-1}\\
~+H_1-(-1)^mH_2+H^* ~~{\rm for} ~mn=2L+1.
\end{array}
\end{array}\eqno{(33)}$$
Here $\Delta_1=\Delta_3=\cdots=\Delta_{mn-1}=1$, which are same with the eigenvalues
of the operators $X_r$ in Ref. [4].
We note that these two results above can be combined due to
$\xi_{-r}=\xi_r$ and $\xi_{mn}=2[H_1-(-1)^mH_2+H^*]$.
So Eqs. (33) have the compact form
$$\begin{array}{l}
\ln\lambda_{\max}-\frac{1}{2}mn\ln[2\sinh(2H)]
        =\frac{1}{2}\sum^{mn}_{r=1}\xi_{2r-1}\\
=\frac{1}{2}\sum^{mn}_{r=1}\cosh^{-1}
       [\cosh(2H_1)\cosh(2H_2)\cosh(2H^*)\\
~~    -\sinh(2H_1)\cosh(2H_2)\sinh(2H^*)\cos\frac{(2r-1)\pi}{2mn}\\
~~    -\cosh(2H_1)\sinh(2H_2)\sinh(2H^*)\cos\frac{(2r-1)\pi}{2n}\\
~~    +\sinh(2H_1)\sinh(2H_2)\cosh(2H^*)\cos\frac{(m-1)(2r-1)\pi}{2mn}].
\end{array}\eqno{(34)}$$

In order to calculate the partition function per atom
$\lambda_{\infty}=\lim_{m,n\rightarrow \infty} (\lambda_{{\rm max}})^{\frac{1}{mn}}$
for the infinite crystal, we replace the sum in Eq. (34) by the integral
$$\ln\lambda_{\infty}=\frac{1}{2}\ln[2\sinh(2H)]
        +\frac{1}{2\pi}\lim_{m\rightarrow \infty}\int^{\pi}_{0}\xi_m(\omega)d\omega,
\eqno{(35)}$$
where
$$\begin{array}{l}
\cosh\xi_m(\omega)={\cal D}(\omega)=\cosh(2H_1)\cosh(2H_2)\cosh(2H^*)\\
~~   -\sinh(2H_1)\cosh(2H_2)\sinh(2H^*)\cos\omega\\
~~   -\cosh(2H_1)\sinh(2H_2)\sinh(2H^*)\cos(m\omega)\\
~~   +\sinh(2H_1)\sinh(2H_2)\cosh(2H^*)\cos[(m-1)\omega].
\end{array}\eqno{(36)}$$
Similarly, the continuous ${\cal A}(\omega)$, ${\cal A}^*(\omega)$,
${\cal A}^\prime(\omega)$, ${\cal B}(\omega)$, ${\cal B}^*(\omega)$,
${\cal B}^\prime(\omega)$, ${\cal C}(\omega)$, $\xi_m(\omega)$,
$\eta(\omega)$, $\eta^*(\omega)$, and $\eta^\prime(\omega)$ replace the discrete
${\cal A}_r$, ${\cal A}^*_r$,
${\cal A}^\prime_r$, ${\cal B}_r$, ${\cal B}^*_r$,
${\cal B}^\prime_r$, ${\cal C}_r$, $\xi_r$,
$\eta_r$, $\eta^*_r$, and $\eta^\prime_r$, respectively,
by letting $\omega=\frac{r\pi}{mn}$.
Here we emphasis that when $H_2=0$, or $H_1=0$,
Eq. (35) is nothing but the Onsager's famous
result in the 2D case [4]. We also note that very different from the 2D case,
the partition function of 3D Ising model is oscillatory with $m$. Therefore,
the conjectured values extrapolating to the infinite system in the numerical calculations
seem to be inaccurate, and the 3D finite-size scaling theory must be modified.

For a crystal of $N=mnl$, the free energy
$$F=U-TS=-Nk_BT\ln\lambda_{\infty},\eqno{(37)}$$
the internal energy
$$\begin{array}{lll}
U&=&F-T\frac{dF}{dT}=Nk_BT^2\frac{\ln\lambda_\infty}{dT}\\
&=&-Nk_BT[H_1\frac{\partial\ln\lambda_\infty}{\partial H_1}
    +H_2\frac{\partial\ln\lambda_\infty}{\partial H_2}
    +H\frac{\partial\ln\lambda_\infty}{\partial H}],
\end{array}\eqno{(38)}$$
and the specific heat
$$\begin{array}{ll}
C&=\frac{dU}{dT}=Nk_B[H^2_1\frac{\partial^2\ln\lambda_\infty}{\partial H^2_1}
       +H^2_2\frac{\partial^2\ln\lambda_\infty}{\partial H^2_2}
       +H^2\frac{\partial^2\ln\lambda_\infty}{\partial H^2}\\
& +2H_1H_2\frac{\partial^2\ln\lambda_\infty}{\partial H_1\partial H_2}
   +2H_1H\frac{\partial^2\ln\lambda_\infty}{\partial H_1\partial H}
   +2H_2H\frac{\partial^2\ln\lambda_\infty}{\partial H_2\partial H}].
\end{array}\eqno{(39)}$$
Here,
$$\begin{array}{l}
\frac{\partial\ln\lambda_\infty}{\partial H_1}=
     \frac{1}{\pi}\lim_{m\rightarrow\infty}\int^{\pi}_{0}
     \cos\eta^*d\omega,\\
\frac{\partial\ln\lambda_\infty}{\partial H_2}=
     \frac{1}{\pi}\lim_{m\rightarrow\infty}\int^{\pi}_{0}
     \frac{\frac{\partial {\cal D}}{\partial H_2}}{\sinh\xi_m}d\omega,\\
\frac{\partial\ln\lambda_\infty}{\partial H}=
     \cosh(2H^*)-
     \frac{1}{\pi}\sinh(2H^*)\lim_{m\rightarrow\infty}\int^{\pi}_{0}
     \cos\eta d\omega,\\
\frac{\partial^2\ln\lambda_\infty}{\partial H^2_1}=
       \frac{2}{\pi}\lim_{m\rightarrow\infty}\int^{\pi}_{0}
     \sin^2\eta^*\coth\xi_m d\omega,\\
\frac{\partial^2\ln\lambda_\infty}{\partial H^2_2}=
       \frac{1}{2\pi}\lim_{m\rightarrow\infty}\int^{\pi}_{0}
       [4-\frac{1}{\sinh^2\xi_m}(\frac{\partial {\cal D}}{H_2})^2]
       \coth\xi_m d\omega,\\
\frac{\partial^2\ln\lambda_\infty}{\partial H^2}=2\sinh^2(2H^*)[
       \frac{1}{\pi}\coth(2H^*)\lim_{m\rightarrow\infty}\int^{\pi}_{0}
        \cos\eta d\omega\\
~~~   +\frac{1}{\pi}\lim_{m\rightarrow\infty}\int^{\pi}_{0}
     \sin^2\eta \coth\xi_m d\omega-1],\\
\frac{\partial^2\ln\lambda_\infty}{\partial H_1\partial H_2}=
     \frac{1}{\pi}\lim_{m\rightarrow\infty}\int^{\pi}_{0}
     \frac{d\omega}{\sinh\xi_m}(\frac{\partial {\cal A}^*}{\partial H_2}
     -\frac{\partial {\cal D}}{\partial H_2}\cos\eta^*\coth\xi_m),\\
\frac{\partial^2\ln\lambda_\infty}{\partial H_1\partial H}=
     -\frac{1}{\pi}\sinh(2H^*)\\
~~~~~     \times\lim_{m\rightarrow\infty}\int^{\pi}_{0}
     \frac{d\omega}{\sinh\xi_m}(\frac{\partial {\cal A}^*}{\partial H^*}
     -2\cosh\xi_m\cos\eta\cos\eta^*),\\
\frac{\partial^2\ln\lambda_\infty}{\partial H_2\partial H}=
     -\frac{1}{\pi}\sinh(2H^*)\\
~~~~~    \times\lim_{m\rightarrow\infty}\int^{\pi}_{0}
     \frac{d\omega}{\sinh\xi_m}(\frac{\partial A}{\partial H_2}
     -\frac{\partial {\cal D}}{\partial H_2}\cos\eta\coth\xi_m).
\end{array}$$
We note that at the critical point, $\lim_{\omega\rightarrow 0}\xi_m
\rightarrow 0$. However, $\lim_{\omega\rightarrow 0}\frac{\partial{\cal D}}
{\partial H_2}/\sinh\xi_m\rightarrow -\cos\eta(0)$. Therefore, we can see
from Eqs. (37) and (38) that at the critical point, the internal energy $U$ is continuous
while the specific heat $C$ becomes infinite, similar to the 2D case.

We consider the special case of $J_1=J_2$, where the calculation of
the thermodynamic functions can be  simplified considerably. After
integrating, Eq. (36) can be rewritten as
$$\begin{array}{l}
\cosh\xi_\infty(\omega)=\cosh(2H_1)\cosh(2H_{2D}^*)\\
~~~~~~~~~~~~~~~~~~   -\sinh(2H_1)\sinh(2H_{2D}^*)\cos\omega,
\end{array}\eqno{(40)}$$
where
$$ H_{2D}^*=H^*-H_1.\eqno{(41)}$$
It is surprising that Eq. (40) is nothing but that in 2D Ising model
with the interaction energies $(J_1,J_{2D})$ and $H_{2D}=\frac{J_{2D}}{k_BT}$.
Therefore, $\ln\lambda_{\infty}-\frac{1}{2}\ln[2\sinh(2H)]$ in three
dimensions can be obtained from $\ln\lambda^{2D}_{\infty}
-\frac{1}{2}\ln[2\sinh(2H_{2D})]$ in two dimensions by taking the
transformation (41). In other words, the thermodynamic properties
of 3D Ising model originate from those in 2D case. We can also see from
Eq. (41) that both 2D and 3D Ising systems approach simultaneously
the critical point, i.e. $H^*_{2D}=H_1$ and $H^*=2H_1$.
It is expected that the scaling laws near the critical point
in two dimensions also hold in three dimensions [6].

The energy $U$ and the specific heat $C$ of 2D Ising model with
the quadratic symmetry (i.e. $H_1=H_{2D}$) have been calculated
analytically by Onsager and can be expressed in terms of
the complete elliptic integrals [4]. The critical exponent associated
with the specific heat $\alpha_{2D}=0$. Because 3D Ising model with
the simple cubic symmetry (i.e. $H_1=H_2=H$) can be mapped exactly
into 2D one by Eq. (41), the expressions of $U$ and $C$ in three dimensions
have similar forms with those in two dimensions. So the critical exponent
$\alpha_{3D}$ of the 3D Ising model is identical to $\alpha_{2D}$, i.e.
$\alpha_{3D}=0$. According to the scaling laws $d\nu=2-\alpha$ and
$\mu+\nu=2-\alpha$ [6], we have $\nu_{3D}=\frac{2}{3}$ and $\mu_{3D}=
\frac{4}{3}$.

Up to now, we have obtained the partition function per site and some physical
quantities when the $z$ axis is chosen as the transfer matrix direction.
However, if the $x(y)$ axis is parallel to the transfer matrix direction,
the corresponding partition function per site can be achieved from Eqs. (35)
and (36) by exchanging the interaction constants along the $x(y)$ and $z$ axes.
Therefore, the total physical quantity in 3D Ising model, such as the free energy,
the internal energy, the specific heat, and etc., can be calculated by taking
the average over three directions. We note that the average
of a physical quantity naturally holds for 2D Ising model.

\section{V. High temperature expansions}

Now we calculate the high temperature expansions of the partition function per atom
when $J_1=J_2=J$. According to the identity
$$\int^{2\pi}_{0}{\rm ln}(2{\rm cosh} x-2{\rm cos}\omega^\prime)d\omega^\prime=2\pi x,
\eqno{(42)}$$
from Eqs. (35) and (36), we obtain
$$\begin{array}{lll}
{\ln}\frac{\lambda_\infty}{2}&=&\frac{1}{2\pi^2}\int^\pi_0\int^\pi_0{\ln}\{{\cosh}^3(2H)\\
 & &-{\sinh}(2H){\cosh}(2H)[{\cos}\omega+{\cos}(m\omega)] \\
 & &+{\sinh}^2(2H){\cosh}(2H){\cos}[(m-1)\omega]\\
 & &-{\sinh}(2H){\cos}\omega^\prime\}d\omega d\omega^\prime \\
 &=&3*{\ln}({\cosh}(H))+\frac{3}{2}{\ln}(1+k^2)\\
 & &+\frac{1}{2\pi^2}\int^\pi_0\int^\pi_0{\ln}\{1-\frac{2k(1-k^2)}{(1+k^2)^2}[{\cos}\omega+{\cos}(m\omega)]\\
 & &+\frac{4k^2}{(1+k^2)^2}{\cos}[(m-1)\omega]\\
 & &-\frac{2k(1-k^2)^2}{(1+k^2)^3}{\cos}\omega^\prime\}d\omega d\omega^\prime \\
 &=&3*{\ln}({\cosh}(H))-3k^4-62k^6-\frac{2081}{2}k^8\\
 & &-21024k^{10}-\cdots,
\end{array}\eqno{(43)}$$
where $k=\tanh H$. Therefore, the partition function per atom in high temperatures is
$$\lambda_\infty=2\cosh^3H (1-3k^4-62k^6-1036k^8-20838k^{10}-\cdots).
\eqno{(44)}$$
We note that for PBCs, the high temperature partition function per atom reads [20]
$$\lambda^p_\infty=2\cosh^3H (1+3k^4+22k^6+192k^8+2046k^{10}+\cdots).
\eqno{(45)}$$
Obviously, the difference between $\lambda_\infty$ and $\lambda^p_\infty$
comes from the screw boundary condition along the ${\bf X}$ direction (see Fig. 1).
We note that the $k^2$ term in Eqs. (44) and (45) vanishes,
which can be seen as a feature of 3D Ising model.

\section{VI. Conclusions}

We have exactly solved 3D Ising model by an algebraic approach.
The critical temperature $T_{ic} (i=1, 2, 3)$, at which an order transition
occurs, is determined. The expression of $T_{ic}$ is consistent with the exact formula
in Ref. [17]. At $T_{ic}$, the internal energy is continuous
while the specific heat diverges. We note that if and only if
the screw boundary condition along the $(1 0 0)$ direction and the periodic boundary conditions
along both $(0 1 0)$ and $(0 0 1)$ directions are imposed, the Onsager operators (15) along {\bf Y}
direction can form a closed Lie algebra, and then the Hamiltonian $H_y$ (16) is obtained rigorously.
For PBCs, the Onsager operators  along {\bf X} or {\bf Y} direction cannot construct a Lie algebra,
and hence 3D Ising model is not solved exactly. Therefore, the numerical simulations on
3D finite Ising model with PBCs are unreliable due to the unclosed spin configurations on the
transfer matrix plane. It is known that the BCs (the surface terms) affect heavily the results on small system,
which lead to the different values extrapolating to the infinite system.
However, the impact of the BCs on the critical temperatures can be neglected in the thermodynamic limit.
Because the partition function per atom of
3D Ising model with $H_1=H_2$ is equivalent to that of a 2D Ising model,
the thermodynamic properties in three dimensions are highly correlated to those of 2D Ising system.
When the interaction energy in the third dimension vanishes,
the Onsager's exact solution of 2D Ising model is recovered immediately.
This guarantees the correctness of the exact solution of 3D Ising model.

\section{ACKNOWLEDGEMENTS}

This work was supported by the Sichuan Normal University and the "Thousand
Talents Program" of Sichuan Province, China.

\end{document}